\documentclass[sigconf,nonacm]{acmart}

\AtBeginDocument{%
  }

\setcopyright{acmlicensed}
\copyrightyear{2018}
\acmYear{2018}
\acmDOI{XXXXXXX.XXXXXXX}
\acmConference[Conference acronym 'XX]{Make sure to enter the correct
  conference title from your rights confirmation email}{June 03--05,
  2018}{Woodstock, NY}
\acmISBN{978-1-4503-XXXX-X/2018/06}

\begin{document}

\title{Beyond Perspectives: A Trio-Ethnography of Interpretation Evolution in LLM-Supported Programming Education}

\author{Jennie Ren}
\email{ren\_c@mercer.edu}
\affiliation{%
  \institution{Mercer University}
  % \city{Macon}
  % \state{Georgia}
  \country{United States}}

\author{Jordan H. McDowell}
\email{Jordan.H.Mcdowell@live.mercer.edu}
\affiliation{%
  \institution{Mercer University}
  % \city{Macon}
  % \state{Georgia}
  \country{United States}}

\author{Kyrie Zhixuan Zhou}
\email{kyrie.zhou@utsa.edu}
\affiliation{%
  \institution{University of Texas at San Antonio}
  % \city{Anon City}
  % \state{Anon State}
  \country{United States}}

\begin{abstract}
Generative AI is reshaping programming education, yet educators often infer students' AI-supported learning from classroom observations alone. This experience report presents a trio-ethnography involving two computing educators with different teaching philosophies and one undergraduate computer science student to examine how these interpretations evolve through dialogue. Across three conversations, the educators reflected on students' AI use, discussed changes to programming pedagogy, and revisited their assumptions after engaging with the student's lived experiences. Rather than simply confirming or contradicting the educators' perspectives, the student's narratives revealed learning processes that were largely invisible in the classroom, prompting both educators to reconsider assumptions about AI use, assessment, transparency, and programming instruction. We argue that trio-ethnography offers a valuable reflective approach for helping computing educators move beyond observable student behaviors toward a richer understanding of AI-supported learning and for informing instructional adaptation in the era of generative AI.
\end{abstract}




\maketitle

\section{Introduction}

Generative AI has rapidly transformed programming education by changing how students learn programming, complete programming assignments, and solve programming problems. As large language models (LLMs) become increasingly integrated into programming learning, they can generate solutions to programming assignments while also explaining concepts, debugging errors, and supporting independent learning. Consequently, observable programming products, such as submitted code or completed assignments, no longer necessarily reflect students' learning processes and outcomes.As a result, a growing gap has emerged between what educators can observe from programming products and how students actually learn with AI, making it increasingly difficult for educators to accurately interpret students' learning in the LLM era.

This gap has important implications for programming education because effective instruction depends on educators' ability to accurately interpret how students learn.Educators make decisions about assessment, feedback, classroom activities, and instructional strategies based on how they interpret students' learning. When these interpretations are inaccurate or incomplete, instructional decisions may become misaligned with students' actual learning needs, resulting in instructional practices that are less effective in supporting student learning. Therefore, there is a need for approaches that help educators develop more accurate interpretations of students' AI-supported learning.

To address this need, we employed trio-ethnography involving two computing educators and one undergraduate computer science student.Rather than using trio-ethnography simply to compare educators' and students' perspectives on AI-supported programming learning, we used it as a process for developing more accurate interpretations of students' learning through dialogue.Interestingly, we found that accurate interpretations did not emerge immediately.Instead, educators' interpretations evolved throughout the dialogue as their initial assumptions were refined, confirmed, or complicated through engagement with the student's lived experiences.This interpretive evolution ultimately enabled educators to develop more accurate understandings of students' AI-supported learning and reconstruct their teaching beliefs.

This study makes both methodological and educational contributions. Methodologically, it demonstrates how trio-ethnography can be used not merely to compare educators' and students' perspectives, but to examine how educators' interpretations evolve through dialogue toward more accurate understandings of students' AI-supported learning. By narrowing the gap between educators' interpretations and students' actual learning processes, this methodological contribution enables an educational contribution by supporting the reconstruction of educators' teaching beliefs based on more accurate interpretations of students' learning. These reconstructed teaching beliefs provide implications for programming education in the LLM era by informing pedagogical reflection and the design of instruction that better supports students' AI-supported learning.

\section{Related Work}

\subsection{Generative AI in Programming Education}

Recent computing education research has examined how generative AI is reshaping programming learning, teaching, and assessment. 
Large language models can generate programming solutions, explanations, exercises~\cite{sarsa2022automatic,becker2023programming}, and responses to help-seeking requests~\cite{hellas2023help}, creating new opportunities for instructional support~\cite{denny2024computing} while also raising concerns about over-reliance and academic integrity~\cite{becker2023programming}.
Studies of novice programmers using tools such as ChatGPT and GitHub Copilot show that students use generative AI to generate code~\cite{kazemitabaar2023effect}, understand syntax, debug errors, and seek explanations~\cite{prather2023s}, and complete programming tasks~\cite{prather2024widening}.
Also highlighted are risks for novice learning, including reduced metacognitive engagement and difficulty evaluating generated code~\cite{prather2024widening}, overconfidence, and uneven benefits across learners~\cite{margulieux2024selfreg}.

Other work has examined how instructors respond to generative AI in computing education. 
Researchers have studied instructor perceptions~\cite{zastudil2023generativeAI,sheard2024instructor,prather2024instructors}, course policies~\cite{lau2023ban,ali2025policies}, AI-assisted assessment~\cite{hoq2024detecting}, AI-based tutoring systems~\cite{zamfirescu2025bot}, prompt-based programming exercises~\cite{denny2024prompt}, and emerging guidelines for responsible classroom integration~\cite{cambaz2024aiCodeGeneration,liu2024cs50,maguire2025declared}. Collectively, this literature shows that generative AI is not merely a code-generation tool but a technology that reshapes programming practice, learning support, assessment design, and classroom policy. 
However, most studies characterize AI tools, student uses, instructor perceptions, or educational risks. 
Less attention has been paid to how educators develop more accurate interpretations of students' AI-supported learning, especially when observable programming products may not reveal the learning processes behind them.

\subsection{Dialogic Approaches to Understanding Educators and Students}

Prior research has used qualitative and dialogic approaches to understand educational experiences, identities, and perspectives.
Duoethnography positions participants as co-inquirers who place their lived experiences in dialogue to question assumptions and generate new understandings~\cite{sawyer2012duoethnography,sawyer2013duoethnography}. 
In education, duoethnography and related reflective approaches have been used to examine professional identity~\cite{long2021duoethnographic,cakcak2024becoming}, language teacher education~\cite{banegas2021critical}, and collaborative reflection~\cite{chang2016collaborative,adams2015autoethnography}. 
Trio-ethnography extends this dialogic tradition by bringing three participants into reflective conversation and has been used in HCI, accessibility~\cite{jain2020trioethnography}, and education research~\cite{desai2023chatgptTrio,le2021trioethnography} to surface multiple lived experiences around technology, disability, professional practice, and learning.

These studies demonstrate the value of dialogic methods for surfacing multiple perspectives and supporting reflective understanding. However, dialogue is primarily used to understand or compare participants' perspectives, such as what educators think, what students experience, and how these perspectives differ. Our study uses trio-ethnography for a different purpose by positioning student dialogue as evidence that helps educators refine and reconstruct their interpretations of students' AI-supported learning. In this way, trio-ethnography becomes a mechanism for narrowing the gap between what educators can observe and how students actually learn with AI.

\section{Method}

\subsection{Research Design}

To investigate how trio-ethnography can help narrow the gap between what educators can observe and how students actually learn with AI, this study employed this approach to examine how computing educators' interpretations of students' AI-supported programming evolved through dialogue and subsequently reshaped their teaching beliefs and instructional practices. Trio-ethnography positions participants as co-inquirers who critically examine, question, and reconstruct one another's interpretations. Because generative AI is rapidly reshaping computing education, we selected trio-ethnography to capture not only how interpretations evolve through dialogue, but also how this process informs educators' teaching beliefs and future instructional decision-making.

\subsection{Participants}

The trio consisted of two computing educators (R1 and R2) and one undergraduate computer science student (R3). R1 is an assistant professor in Computer Science with experience teaching undergraduate courses, including programming, web development, Java, and mobile application development. R1 initially expressed significant concern about students' over-reliance on AI-generated code. R2 is an assistant professor in Information Systems who teaches introductory programming to undergraduate business students and has adopted an open approach toward integrating generative AI into programming instruction. R3 is an undergraduate Computer Science student who regularly uses generative AI while learning programming. The student joined the dialogue as an equal contributor whose lived experiences extended and challenged the educators' interpretations. The trio was intentionally composed of two educators with contrasting instructional perspectives and one student who regularly used generative AI, enabling educators to examine and reflect on their interpretations through direct engagement with the student's perspectives.

\subsection{Procedure}

The study was conducted in three analytical stages.

\noindent\textbf{Stage 1: Initial Educator Dialogue.}
The two computing educators first engaged in a series of duo-ethnographic discussions guided by three reflection topics: (1) how generative AI has changed students' programming learning, (2) how computing educators should adapt programming instruction in response to these changes, and (3) how these changes have influenced their teaching beliefs. Across these discussions, the educators articulated and developed their initial interpretations of students' AI-supported learning based on their classroom experiences. As they questioned and reflected on one another's perspectives, some of these interpretations were further refined through the dialogue itself.

\noindent\textbf{Stage 2: Student Dialogue.}
An undergraduate computer science student was subsequently interviewed using a semi-structured protocol designed to explore the student's lived experiences of learning programming with generative AI. Instead of mirroring the educators' discussion prompts, the interview questions were developed to elicit student perspectives that could respond to, elaborate on, and challenge the educators' initial interpretations. The student's narratives confirmed, extended, and complicated the educators' initial interpretations of AI-supported learning.

\noindent\textbf{Stage 3: Educator Reflective Reconstruction.}
Following the student dialogue, the educators revisited their original discussions, treating the student's dialogue as evidence to refine and reconstruct their interpretations of students' AI-supported learning. This reflective process reconstructed the educators' interpretations of students' AI use,  enabling more accurate understandings of students' AI-supported learning,  refined their teaching beliefs, and informed subsequent instructional practices.

\subsection{Data Analysis}
The conversations from Stage 1 (Initial Educator Dialogue) and Stage 2 (Student Dialogue) served as the primary data sources for analysis. These conversations were audio-recorded, transcribed using Otter.ai, and manually reviewed for accuracy.

Through multiple rounds of transcript reading and collaborative discussion, we examined how educators' interpretations of students' AI-supported programming learning were confirmed, extended, challenged, or complicated by the student's lived experiences. We also paid particular attention to recurring educator interpretations, as these reflected deeply held assumptions and often became the most meaningful sites of interpretive reconstruction.

Rather than viewing these interactions as isolated observations, we treated them as evidence of an evolving interpretive process. By comparing the educators' initial interpretations with the student's perspectives, we reflected on how these interpretations were reconstructed through dialogue. This reconstruction subsequently refined the educators' teaching beliefs and informed future instructional practices in programming education.

\section{Results}

Following data analysis, the findings reveal an evolving interpretive process through which educators gained increasingly accurate understandings of students' AI-supported programming learning through dialogue with the undergraduate student. The findings illustrate how the student's lived experiences prompted educators to re-examine their initial interpretations and reshape their teaching beliefs.

\subsection{From AI Stigma to AI Usage Transparency to Explicit AI Guidance}
The educators' interpretations of students' AI use evolved from identifying an emerging \emph{AI stigma}, to recognizing the importance of AI usage transparency, and ultimately to emphasizing explicit guidance for appropriate AI use. These interpretations, initially developed through educator dialogue, were later confirmed by the student's experiences and reconstructed into new teaching beliefs.

The dialogue began with R1 describing students who submitted programming assignments containing syntax beyond what had been taught in class while claiming they had not used AI.

\begin{quote}
\textit{``Some students used syntax that I never taught in class... but in their reflection they still said they didn't use AI.''}
\end{quote}

Initially, R1 interpreted this behavior as students deliberately concealing AI use because they believed it was academically inappropriate.

\begin{quote}
\textit{``I think the students feel shameful about using AI... probably they think the instructor does not allow us to use AI.''}
\end{quote}

Rather than viewing nondisclosure simply as dishonesty, the educators interpreted it as evidence of an emerging \emph{AI stigma}, in which students perceived AI use as something hidden from instructors.

The dialogue then shifted toward instructors' responsibility. R2 questioned whether expectations about AI use had ever been explicitly communicated, prompting R1 to reflect:

\begin{quote}
\textit{``I didn't say anything. I didn't teach them what should be a good way to use AI, or what should be a bad way to use AI.''}
\end{quote}

Through this reflection, the educators' interpretation shifted from viewing AI stigma as a student issue to recognizing that it was also shaped by instructional ambiguity. Rather than assuming students understood classroom expectations, they recognized that instructors needed to communicate those expectations explicitly. This refined interpretation led R1 to conclude:

\begin{quote}
\textit{``I think AI usage transparency is very important.''}
\end{quote}

Here, AI usage transparency extended beyond encouraging disclosure. Instead, it meant making instructors' expectations visible so students understood when AI use was appropriate and how it could support learning.

Comparing their teaching practices further refined this interpretation. Unlike R1, R2 intentionally discussed AI expectations during the first class meeting.

\begin{quote}
\textit{``I spent maybe ten minutes in the first class telling students what is allowed and what is not allowed with AI use.''}
\end{quote}

This comparison led both educators to recognize that transparency alone was insufficient. Students also needed explicit guidance on how AI should support programming learning. Their interpretation therefore evolved beyond transparency toward \emph{explicit AI guidance}, emphasizing that instructors should teach productive AI-supported learning strategies rather than simply establish classroom rules.

The subsequent student dialogue confirmed this refinement. The student described feeling uncertain about AI use because AI-assisted programming had been prohibited in high school. The student also expressed a preference for instructors to communicate expectations explicitly and, more importantly, teach students how AI could be used effectively for learning rather than simply whether it was permitted.

Reflecting across both educator and student dialogues, the educators reconstructed their understanding of AI-supported programming instruction. Their reconstructed teaching beliefs emphasized that AI should become an explicit, discussable, and teachable component of programming education through transparent expectations and purposeful guidance that enables students to use AI to support, rather than replace, their learning.

\subsection{Redefining AI's Role: From Answer Provider to Learning Partner and Tutor}

A second interpretive thread centered on how the educators understood AI's educational role in programming learning. Although both educators initially expressed concern that students were treating AI as an \emph{answer provider} that replaced independent thinking, their dialogue refined this interpretation by recognizing that AI itself was not the educational problem. Instead, the educators came to view AI as a \emph{learning partner}, and more specifically, as a \emph{tutor} that could scaffold students' conceptual understanding while preserving meaningful learning. This refined interpretation was subsequently confirmed by the student's lived experiences and reinforced through educator reflection, 
positioning AI as a pedagogical learning partner in programming education.

Initially, R1's interpretation was shaped by observations that many students relied on AI to obtain programming solutions rather than understand programming concepts.

\begin{quote}
\textit{``I feel like the students just got used to copy and paste the code... they're not using AI to help them, they're just using AI like, `I can get the answer from AI,' and just paste the answer.''}
\end{quote}

Rather than criticizing AI itself, R1's concern centered on its educational role. When AI functioned as an \emph{answer bank}, students risked replacing their own reasoning with AI-generated solutions, reducing opportunities for conceptual learning.

The dialogue gradually refined this interpretation. Rather than debating whether AI should be permitted, R2 encouraged reconsidering how AI could support learning. Drawing on software engineering practice and his teaching experience, he argued that AI-generated code was not inherently problematic; students' learning depended on how they engaged with it.

\begin{quote}
\textit{``I don't really care if students use AI to generate the initial version of the code... if they have prior experience with programming by themselves, I don't feel like AI coding is a major concern.''}
\end{quote}

This discussion shifted the educators' interpretation from viewing AI as an answer provider toward recognizing it as a learning partner. More specifically, AI was increasingly conceptualized as a tutor that helps students understand programming concepts, explain unfamiliar syntax, and support problem solving while leaving learning in students' hands.

The student dialogue strongly confirmed this refined interpretation. Rather than describing AI as a tool for obtaining complete solutions, the student consistently portrayed it as a learning resource for understanding Java syntax, programming concepts, and code.

\begin{quote}
\textit{``It was really helping me out... I was studying the syntax... Now you can learn the syntax.''}
\end{quote}

Throughout the interview, the student described interacting with AI by asking questions, seeking explanations, and improving conceptual understanding rather than simply requesting complete answers. These narratives aligned with the educators' refined interpretation that AI could function as a tutor supporting students' learning.

Reflecting across both educator and student dialogues, the educators reconstructed AI's educational role. Their reconstructed teaching beliefs emphasized positioning AI as a learning partner and tutor that supports conceptual understanding, reasoning, and independent learning.

\subsection{Invisible Learning Beyond AI-Generated Answers}

A third interpretive thread centered on educators' persistent interpretation that students often copied AI-generated answers without meaningful learning. Unlike the previous themes, this interpretation was \emph{complicated} rather than confirmed by the student's experiences. During the educator dialogue, R1 refined this concern through the idea of ``one more step'': AI could support learning if students continued thinking after receiving AI-generated code. The student later complicated this refined interpretation by revealing multiple learning activities that occurred after AI-generated responses—activities that often remained invisible to instructors.

Initially, R1 interpreted students' AI use as obtaining answers rather than learning programming. From this perspective, AI-generated code appeared to replace students' own reasoning.

\begin{quote}
\textit{``I feel like the student, they just got used to copy paste the code... they're not using the AI to help them, they're just using the AI, like, oh, I can get the answer from the AI, and just paste the answer.''}
\end{quote}

This interpretation was grounded in classroom evidence. R1 could observe students' submitted code, including syntax or programming techniques beyond what had been taught. However, these observable products provided limited access to what students had actually done before submission. As a result, AI-generated or unusually advanced code was often interpreted as evidence of copy-and-paste behavior rather than meaningful learning.

As the educator dialogue unfolded, this interpretation was refined. R2 suggested that AI-generated code was not inherently problematic if students continued engaging with it. R1 therefore reframed the issue as whether students took an additional step after receiving AI-generated solutions.

\begin{quote}
\textit{``If they have a further step... they use the code generated by AI, then they can step on this kind of thing to have the further thinking, that's fine, but if they just... copy paste the code, that's not fine.''}
\end{quote}

This idea of ``one more step'' became a turning point in the educators' interpretation. The educators distinguished between AI use that ended with answer generation and AI use that initiated further reasoning. However, they still had limited evidence regarding whether students actually engaged in these further learning steps.

The subsequent student dialogue complicated this refined interpretation. The student confirmed that AI could reduce the effort required to obtain a solution, but also described additional learning activities after receiving AI-generated help.

\begin{quote}
\textit{``Usually I have notes to take down, I practice, and I will just create more of what I was trying to solve the first time, and just make sure I'm able to do this from muscle memory.''}
\end{quote}

The student's narrative expanded the educators' idea of ``one more step'' into multiple learning steps. After receiving AI-generated assistance, the student described reading code, analyzing its logic, taking notes, practicing independently, modifying solutions, and verifying understanding. These activities suggested that AI interaction did not always end with copying a final answer; AI-generated responses often became the beginning of a longer learning process that remained largely invisible to instructors.

These accounts complicated the educators' persistent interpretation without fully invalidating it. Some students may indeed copy AI-generated code without reflection. However, the student narratives revealed that similar programming submissions could result from different learning trajectories. One student might submit code after copying an AI-generated answer, while another might submit a similar product after reading, questioning, modifying, practicing, and verifying understanding. From the instructor's perspective, both submissions may appear similar, but the learning processes behind them may differ substantially.

Reflecting across both educator and student dialogues, the educators reconstructed their understanding of AI-supported learning. Their reconstructed teaching beliefs emphasized recognizing the invisible learning processes that may occur beyond AI-generated answers.

\subsection{Reconstructing Teaching Beliefs Through Student Dialogue}

The final interpretive thread centered on how student dialogue reshaped the educators' teaching beliefs in the LLM era. As a result, the educators began to reimagine assessment, debugging instruction, and active learning opportunities.

%Initially, both educators believed that meaningful programming learning required students to actively construct understanding through practice rather than passively receive or reproduce code. R1 summarized this belief by reflecting that programming learning required not only doing, but actively doing.

%\begin{quote}
%\textit{``Right now I just add on one word: actively doing something is the only way to learn something.''}
%\end{quote}

The student dialogue revealed reasoning processes that required greater instructional support in AI-supported programming learning. Although the student described AI as helpful for learning syntax and programming concepts, the dialogue also revealed situations where AI produced a working solution without helping the student understand why it worked. When comparing personal code with AI-generated code, the student explained that the two versions appeared similar, yet only the AI-generated version worked.

\begin{quote}
\textit{``What I had versus what the AI had... it was pretty similar, but mine was incorrect. While the AI's was correct, I wasn't really sure why and how that was different.''}
\end{quote}

This student perspective prompted the educators to reconsider what programming instruction needed to make visible. The educators recognized the importance of helping students reason about why code worked, why errors occurred, and how AI-generated solutions differed from their own attempts.

The student dialogue reshaped the educators' beliefs about debugging instruction. The student's difficulty in understanding why AI-generated code worked while personal code failed suggested that students needed explicit support in comparing code, interpreting errors, and reasoning through debugging processes. During the educator dialogue, R1 similarly reflected that debugging instruction should focus not only on fixing errors but also on understanding the reasoning behind errors.

\begin{quote}
\textit{``Instead of only focusing on fixing errors... they also need to understand the source, the reason behind the errors.''}
\end{quote}

This reconstructed teaching belief positioned AI as a tool to support students' debugging reasoning by helping them explain errors, compare alternative solutions, and understand why corrections work.

The educators also reconsidered assessment. When students could use AI to quickly generate correct code, completed assignments no longer necessarily represented students' learning outcomes. This prompted the educators to reconsider assessment designs that evaluate students' reasoning in addition to final code, such as monitored assessments or assignments requiring students to explain, justify, or reflect on their solutions. Such opportunities could also make the invisible learning processes identified in Theme 3 more visible to educators.

Finally, the educators' reconstructed teaching beliefs emphasized active learning opportunities that reveal students' reasoning. Across the dialogue, both educators increasingly valued practices such as live coding, active recall, and code-output prediction. R1 described future instruction as creating more opportunities for students to actively learn rather than passively follow teacher-guided demonstrations.

\begin{quote}
\textit{``As a teacher, I mean to, in the future, I mean to design more this kind of opportunities for the students, so they can learn.''}
\end{quote}

Reflecting across both educator and student dialogues, the educators reconstructed their teaching beliefs to emphasize instruction that reveals, supports, and develops students' reasoning. These reconstructed teaching beliefs emphasized programming instruction that assesses reasoning, makes learning processes visible, teaches debugging explicitly, and creates active learning opportunities that support learning with AI.

\section{Discussion}

\subsection{Implications for Programming Education}

%\subsubsection{From Understanding Students to Rethinking Programming Education} actually below is a opening paragraph it's not a specific implication, so i delete this subsubsection title, it's just a opening 

Across the trio-ethnographic dialogue, educators' interpretations of students' AI-supported programming learning evolved rather than remained fixed. Student dialogue confirmed key educator interpretations, such as AI transparency and AI as a learning partner, while also complicating persistent assumptions by revealing learning processes that were largely invisible from classroom observations.

These findings shift the focus from judging whether students use AI to understanding how they learn with AI. Observable evidence, such as submitted code or completed assignments, provides only partial access to students' reasoning and learning processes. Incorporating students' perspectives therefore enables educators to develop more accurate understanding of AI-supported programming learning.

More importantly, the trio-ethnographic dialogue extended beyond understanding students to reshaping educators' teaching beliefs. These reconstructed teaching beliefs, in turn, prompted educators to reconsider their programming instruction. The following implications illustrate how this reconsideration informed broader pedagogical reflection in the LLM era.

\subsubsection{Repositioning AI in Programming Education}

A key implication of this study is the need to reposition AI in programming education from an issue of classroom policy to a component of pedagogy.

The reconstructed teaching beliefs point toward two complementary pedagogical directions. First, AI should become a visible, legitimate, and teachable part of programming classrooms. Educators should explicitly communicate when AI use is appropriate, discuss responsible AI use, and provide guidance on how AI can effectively support learning. Second, programming instruction should position AI as a learning partner and tutor that supports students' conceptual understanding, problem-solving skills, and independent reasoning.

Collectively, these implications shift programming education from regulating AI use to designing learning environments where students learn with AI.

% To be written next.

\subsubsection{Making Learning Processes Visible}

A third implication concerns making students' AI-supported learning processes more visible. The findings showed that instructors may observe students' final code or assignment submissions, but these products provide limited evidence of what students actually did after receiving AI-generated assistance. Meaningful learning may occur through reading, analyzing, practicing, modifying, and verifying AI-generated responses, yet these processes often remain invisible in the final programming product.

For programming education, instructors should design activities that make students' reasoning and learning processes more visible. Reflection prompts and process documentation can ask students to explain how they used AI, what they accepted or rejected, what they modified, and what they learned. 

Such practices shift attention from detecting AI use to understanding how students learn with AI, allowing educators to better distinguish between copy-and-paste submission and meaningful AI-supported learning.

% To be written next.

\subsubsection{Methodological Implications of Trio-ethnography}

Beyond its implications for programming education, this study highlights the methodological value of trio-ethnography as a way to examine how educational interpretations evolve through dialogue, rather than merely collecting different perspectives from educators and a student.

This suggests that trio-ethnography moves beyond comparing perspectives by using student dialogue as evidence to progressively refine educators' interpretations and develop more accurate understandings of student learning. This methodological contribution ultimately informs pedagogical reflection and programming education in the LLM era.

% To be written next.

\subsection{Limitations and Future Work}

This study has several limitations that should be considered when interpreting the findings.

First, the trio-ethnography involved only one undergraduate student participant. Although the student's perspectives provided rich insights into AI-supported programming learning, the participant represented a highly motivated learner. Consequently, the student learning strategies described throughout the dialogue, such as reading AI-generated code, taking notes, practicing independently, and verifying understanding, may not represent the experiences of all programming students. Future research should therefore include students with more diverse academic backgrounds, learning motivations, and programming abilities to examine whether similar interpretive patterns emerge across different learner populations.

The study also focused on a single trio-ethnographic dialogue involving two programming educators and one undergraduate student. The reconstructed interpretations and teaching beliefs should therefore be understood as contextually situated rather than universally representative. Future studies could conduct trio-ethnographies across different institutions, programming courses, and educational contexts to examine how educators' interpretations evolve in diverse learning environments.

% To be written after implications.

\section{Conclusion}

This study demonstrates the methodological value of trio-ethnography in computing education. By using student dialogue as evidence to progressively refine educators' interpretations, trio-ethnography supported more accurate understandings of students' AI-supported programming learning.

These reconstructed understandings reshaped educators' teaching beliefs and informed pedagogical reflection for programming education in the LLM era. We hope this work encourages future computing education research to further explore the methodological potential of trio-ethnography.

% \begin{acks}
% \end{acks}

\section{Ethics and Privacy Statement}

This experience report presents a trio-ethnographic inquiry among the authors. All conversations involving the educators' and student's reflections were conducted voluntarily with informed consent.

To protect privacy, pseudonyms (R1, R2, and R3) are used throughout the paper. The findings reflect the participants' individual experiences rather than universal practices and are intended to support transparent, responsible, and equitable integration of generative AI in computing education.

\clearpage
\bibliographystyle{ACM-Reference-Format}
\bibliography{sample-base}

% \appendix









\end{document}